# $\chi^2$ TESTING OF OPTIMAL FILTERS FOR GRAVITATIONAL WAVE SIGNALS: AN EXPERIMENTAL IMPLEMENTATION

L. Baggio and M. Cerdonio,
Dipartimento di Fisica, Università di Padova and INFN, Sezione di Padova, via Marzolo 8, 35100 Padova, Italy.

A. Ortolan and G. Vedovato
INFN, Laboratori Nazionali di Legnaro, via Romea 4, 35020 Padova, Italy.

L. Taffarello and J-P. Zendri
INFN, Sezione di Padova, via Marzolo, 8, 35100 Padova, Italy.

M. Bonaldi, P. Falferi
CeFSA, Centro ITC-CNR, Trento and INFN, Gruppo Collegato di Trento, 38050 Povo, Trento, Italy.

V. Martinucci, R. Mezzena, G.A. Prodi and S. Vitale
Dipartimento di Fisica, Università di Trento and INFN, Gruppo Collegato di Trento, 38050 Povo, Trento, Italy.

**Abstract**

We have implemented likelihood testing of the performance of an optimal filter within the online analysis of AURIGA, a sub-Kelvin resonant-bar gravitational wave detector. We demonstrate the effectiveness of this technique in discriminating between impulsive mechanical excitations of the resonant-bar and other spurious excitations. This technique also ensures the accuracy of the estimated parameters such as the signal-to-noise ratio. The efficiency of the technique to deal with non-stationary noise and its application to data from a network of detectors are also discussed.

## 1    Introduction

Wiener-Kolmogoroff (WK) optimal filter is the main tool of signal extraction for gravitational wave (GW) detectors. In gaussian noise, WK filtering is fully equivalent to maximum likelihood fitting of a signal model to the data. As a consequence, hypothesis testing can be applied to the filter by means of a proper sufficient statistics, as is the case for any other maximum likelihood fit. We have shown recently[1] that, in presence of pure gaussian noise, a likelihood hypothesis test leads to a standard $\chi^2$-test of the "goodness of the fit". These ideas have been implemented within the data analysis of the AURIGA ultra-cryogenic detector[2].

We performed a preliminary bench test of our filtering and event discrimination algorithms by using a room temperature resonant-bar detector. We then applied the algorithms to the AURIGA detector as soon as it started taking data in June 1997. In this paper, we report on the performance of the method and on the procedures we use to cope with the problem of the noise being non-stationary and non-gaussian.

The paper is organized as follows. Sec. 2 we summarize the theory of $\chi^2$-test in the framework of WK filtering theory. In Sec. 3 we draw our model for the detector transfer function and noise spectrum. The experimental setup both for the room temperature test facility and for the cryogenic detector is reviewed in Sec. 4. Sec 5 is devoted to the practical implementation of the WK filter, and results of $\chi^2$ event characterization are reported in Sec. 6. Finally, in Sec. 7 we discuss on the relevance of this technique to the case of a single detector and of a network of GW detectors.

## 2  Signal analysis and hypothesis testing

A simplified model for a GW detector is that of a linear system with an output noise n(t), which is commonly described as a stationary stochastic process with gaussian statistics. In the following of this section we adopt a discrete time domain representation, that is we substitute for n(t), a finite length sequence of samples $n_i \equiv n(i\Delta t)$. In this way we get a set $\{n_i\}$ of gaussian random variables (GRV), with $0 \leq i \leq N$.

If a signal enters the system at time $t_0$, the sampled output of the detector is $x_i = A u_i(t_0, \vartheta_j) + n_i$, where $\{u_i\}$ is the properly normalized signal template, A its amplitude and $\{\vartheta_j\}$ any other parameter set the signal may depend.

A well established result[3] of signal analysis states that the minimum variance, unbiased linear estimate of the amplitude A is the GRV :

$$\hat{A} = \frac{\sum_{ij} \mu_{ij} u_i x_j}{\sum_{hk} \mu_{hk} u_h u_k} \equiv \sigma_{\hat{A}}^2 \cdot \sum_{ij} \mu_{ij} u_i x_j \equiv \sum_j w_j x_j \qquad (1)$$

if one assumes to know u, $t_0$, $\vartheta_l$ and the inverse cross correlation function $\mu_{ij}$ of the noise. Here the $w_j$'s are then the coefficients of the WK filter matched to the signal $\{u_i\}$ and $\sigma_{\hat{A}}^2$ is the variance of $\hat{A}$

$$\sigma_{\hat{A}}^2 = \frac{1}{\sum_{hk} \mu_{hk} u_h u_k} \qquad (2)$$

When the noise has gaussian statistics, the WK filter happens to be also a maximum likelihood estimator. In fact, the likelihood function associated with the data set $\{x_i\}$ is

$$\Lambda(x_1, x_2, \ldots ; A) \propto \exp\left[-\tfrac{1}{2} \sum_{ij} \mu_{ij} (x_i - A u_i)(x_j - A u_j)\right], \qquad (3)$$

where the sum runs over the number of data N. It is straightforward to verify that $\Lambda$ reaches its maximum value for A equal to the value given by eq. (1).

For this same value, the log-likelihood ratio:

$$X \equiv \sum_{ij} \mu_{ij} (x_i - \hat{A} u_i)(x_j - \hat{A} u_j) \qquad (4)$$

reaches a minimum.

It can be easily shown that X in eq. ( 4 ) is a random variable with a standard $\chi^2$ statistics. By performing the transformation $\{y_i \equiv \sum L_{ij} x_j\}$, where L is the *whitening* filter that diagonalize $\mu_{ij}$ ($L^{-1}_{mi} \mu_{mn} L^{-1}_{nj} = \delta_{ij}$), one gets:

$$X \equiv \sum_i (y_i - \hat{A} v_i)^2 \qquad (5)$$

with $\{v_i \equiv \sum L_{ij} u_j\}$. This is the linear least square sum for a standard fit of the function $\{v_i\}$ to the data $\{y_i\}$ which is well known to be $\chi^2$ distributed with N-1 degrees of freedom and to be independent of $\hat{A}$.

In order to evaluate $X$, it is easier to work with the equivalent expression[1]:

$$X = \left[ \sum_{ij} m_{ij} x_i x_j - \frac{\hat{A}^2}{s^2_{\hat{A}}} \right] = \left[ \sum_i y_i^2 - \frac{\hat{A}^2}{s^2_{\hat{A}}} \right] \qquad (6)$$

Eq. ( 6 ) shows that if the n parameters $t_0$ and $\{\vartheta_i\}$ are also unknown, their maximum likelihood estimate is the one that makes $\hat{A}^2/s^2_{\hat{A}}$ a maximum. This is in general a non linear fit, and the resulting X is distributed as a $\chi^2$ with N-n-1 degrees of freedom but only within a linear approximation.

We will use in the following mostly the reduced experimental $c_a^2 \equiv \frac{1}{N-n-1} X$ which is expected to be distributed as a reduced chi-square $c_r^2$. This statistic has unitary mean value for any number of degrees of freedom N-n-1.

The $c_a^2$ can be used as a statistical test of goodness-of-the-fit. It can be used to test for consistency of a priori hypothesis on the signal template $\{u_i\}$, with probability thresholds given either by theoretical predictions or by Monte-Carlo simulations for the non-linear case.

It is worth pointing out that, if a set of data fails the test, the resulting estimates for the amplitude $\hat{A}$ and for the other parameters are, in principle, biased. In this sense, the test appears as an unavoidable step of the overall filtering procedure. The relation between the bias on the amplitude estimate and the value of X can be determined analitically. We already pointed out[4] that, if data contain a signal $\{f_i\}$ different from that $\{v_i\}$ to whom the filter has been matched, the experimental value of X fluctuates around a mean value proportional to the square of the signal-to-noise-ratio $SNR \equiv \hat{A}/s_{\hat{A}}$. In the following of this section we will work on the whitened data, and the signal $\{f_i\}$ and the template $\{v_i\}$ are referred at the output of the whitening filter. The apparent chi-square statistics is:

$$c_a^2 = c_r^2 + \frac{1}{N-n-1} \left\{ (SNR_o^f)^2 - (SNR_o^v)^2 + 2(SNR_o^f SNR_n^f - SNR_o^v SNR_n^v) \right\} \qquad (7)$$

where $SNR_o^f$ is the mean value of the signal to noise ratio for the signal $\{f_i\}$ with a filter matched to it, and $SNR_o^v$ is that with the filter matched to $\{v_i\}$. $SNR_n^f$ and $SNR_n^v$ are their fluctuating parts, i.e. two gaussian random variable with zero mean value and unit variance.

The mean value of $\chi_a^2$ is then:

$$\qquad (8)$$

$$\langle c_a^2 \rangle = \langle c_r^2 \rangle + \frac{1}{N-n-1}\langle (SNR_o^f)^2 - (SNR_o^v)^2 \rangle = 1 + \lambda (SNR_o^v)^2$$

where

$$\lambda = \frac{\sum_{i=1}^{N} f_i^2 \sum_{i=1}^{N} v_i^2 - \left(\sum_{i=1}^{N} f_i v_i\right)^2}{(N-n-1)\left(\sum_{i=1}^{N} f_i v_i\right)^2} \qquad (9)$$

is a value that reduces to zero if $f_i = v_i$. Notice that $\lambda$ is proportional to the square of the bias on the signal-to-noise ratio due to the filter inaccuracy.

## 3  WK filter for modeled GW resonant detectors

A resonant-bar detector coupled to a capacitive electromechanical transducer can be quite accurately modeled by an equivalent lumped elements electrical circuit[5]. It is easy to show that the transfer matrix between any port within the circuit and the readout port, always contains the same series of M poles, the $k^{th}$ pair of complex conjugate poles corresponding to a normal oscillation mode with frequency $\omega_k$ and quality factor $Q_k$. The noise generated by any generator within the circuit is transferred to the output through one of these transfer matrices. The total output noise results from the sum of these contributions plus the wide band noise $S_o$ of the final amplifier. It is easy to calculate that, with these assumptions, the total output noise spectral density is:

$$S(\omega) = S_0 \prod_{k=1}^{M} \frac{(i\omega - q_k)(i\omega + q_k)(i\omega - q_k^*)(i\omega + q_k^*)}{(i\omega - p_k)(i\omega + p_k)(i\omega - p_k^*)(i\omega + p_k^*)}, \qquad (10)$$

where $p_k = i\omega_k - \omega_k/(2Q_k)$ and where the complex zeros $q_k$'s are related to the optimal bandpass WK filter.

$S(\omega)$ possesses a few key features. Firstly, the degrees of the polynomials appearing in the numerator and in the denominator are equal, a consequence of having modeled the wide band noise $S_o$ as purely white. Secondly, poles and zeros appear in pairs $\pm p_k$ and $\pm q_k$, as the noise spectral densities are transferred through the square modulus of transfer functions. Finally, as already mentioned, for each pole (or zero) its complex conjugate also appears, a consequence of reality of circuit elements.

The transfer function for an input GW delta pulse will contain the same poles $\{p_k\}$. Reality imposes then that the output signal $u_\delta(t)$ has a Fourier transform $\tilde{u}_\delta(\omega)$ given by:

$$\tilde{u}_\delta(\omega) = \frac{\prod_{j=1}^{\tilde{M}}(i\omega - r_j)(i\omega - r_j^*)}{\prod_{k=1}^{M}(i\omega - p_k)(i\omega - p_k^*)} \qquad (11)$$

with $M > \tilde{M}$ because of the stability of the system and where the coefficients $r$ are obviously the zeroes of the function.

It is well known that the continuous version of the WK filter function, w(t), has a Fourier transform $\tilde{w}(\omega) = \sigma_{\hat{A}}^2 S^{-1}(\omega) \tilde{u}_k^*(\omega)$. By using eq. (10) and (11) one gets for $\tilde{w}(\omega)$:

$$\tilde{w}(\omega) = \sigma_{\hat{A}}^2 S_0^{-1} \frac{\prod_{j=1}^{\tilde{M}} (i\omega + r_j)(i\omega + r_j^*)}{\prod_{k=1}^{M} (i\omega + q_k)(i\omega + q_k^*)} \prod_k \frac{(i\omega - p_k)(i\omega - p_k^*)}{(i\omega - q_k)(i\omega - q_k^*)}. \quad (12)$$

The WK filter splits up in the product $L(\omega) \cdot M(\omega)$, where

$$L(\omega) = S_0^{-1/2} \prod_k \frac{(i\omega - p_k)(i\omega - p_k^*)}{(i\omega - q_k)(i\omega - q_k^*)} \quad (13)$$

is the *whitening filter* for the noise with PSD $S(w)$. This means that a filter with transfer function $L(\omega)$ produces at its output a noise with spectral density $S_w=1$ when fed at the input with the detector noise with PSD $S(w)$, because $|L(\omega)|^2 S(w)=1$.

$M(\omega)$ is defined by

$$M(\omega) = \sigma_{\hat{A}}^2 S_0^{-1/2} \frac{\prod_{j=1}^{\tilde{M}} (i\omega + r_j)(i\omega + r_j^*)}{\prod_{k=1}^{M} (i\omega + q_k)(i\omega + q_k^*)} \quad (14)$$

It is a bandpass filter around the frequencies $\omega_k \equiv |\text{Im}(q_k)|$ with bandwidths $\Delta\omega_k^{opt} \equiv 2|\text{Re}(q_k)|$ which are usually much larger than $\Delta\omega_k \equiv 2|\text{Re}(p_k)| = \omega_k/Q_k$.

Most of the information needed to process the data is contained within this filter matched to a delta-shaped pulse. The response of the system to any other input signal h(t) can always be written as the time convolution $u_\delta * h$, so that, once data have been filtered with the optimum filter matched to $u_\delta$, one can perform the complete WK filtering for h(t) by a simple convolution of h(t) with the filtered data.

In addition, it turns out that for resonant detectors most of the expected signals have Fourier transforms that are rather flat across the comparatively small post filtering bandwidths (~1 to 10Hz) of the detectors, and are thus indistinguishable from a delta pulse[6].

One can show that, for the case of a resonant-bar detector with resonant transducers, in eq. (14), $M=2$, $\tilde{M}=1$ and $r_1=0$. As a consequence, the band-pass filter $M(\omega)$, which is purely anti-causal, introduces an anti-causal component in the response and cannot be implemented in real time.

In addition, the sets $\{p_k\}$ and $\{q_k\}$, along with $S_0$, are the only relevant parameters that enter both the noise spectrum and the transfer function of the system. As a consequence, a check that the PSD of the whitened data is indeed flat within its statistical error, becomes a very useful consistency test for the accuracy of the filter. In Sec. 5 we show how we feed back the deviation from a white spectrum to an automatic adaptive procedure that updates the values of filter parameters.

**4   Experimental layout**

The AURIGA detector[2] consists of a 2.3tons, 3 meters long Al5056 bar equipped with a capacitive electromechanical transducer and a dc-SQUID preamplifier. The bar hangs on a multiple stage pendulum attenuation system (-240db at 1kHz), kept at 0.2K by a $^3$He-$^4$He dilution refrigerator.

The signal is acquired by an ADC at 4.9kHz and synchronized to UTC by means of a GPS clock[7] (see Fig. 1 ).

The room temperature detector used for some of the test shares almost all the relevant features with the cryogenic detector. The most noticeable difference, besides the absence of the cooling system, is that the voltage across the capacitive transducer is fed to very low noise FET preamplifier.

As far as signal analysis is concerned, the most relevant differences among the two detectors are the Q factors ($\approx 10^4$ for the room temperature detector and slightly larger than $10^6$ for AURIGA) and the post filtering bandwidth $\Delta\omega_k^{opt}$ (corresponding to $\approx 10$ Hz for the room temperature detector and $\approx 1$ Hz for AURIGA).

The room temperature detector mounts an electromechanical capacitive actuator, a de-tuned version of the transducer, placed on the face opposite to the one used to extract the signal. It provides a way to excite the bar with short mechanical bursts that mimic a GW signal.

In order to test the $\chi^2$ performance for spurious excitations due to electrical disturbances, short current pulses were injected into a coil inductively coupled to the amplifier-input leads. These pulses were also used to trigger data acquisition as described in Ref. 8.

## 5  Data analysis and $\chi^2$ evaluation.

The on-line analysis of the AURIGA detector has been described elsewhere[9]. Here we summarize its most relevant features and some of the new elements that are of relevance for the present work. A detailed report on the performance of these new features have also been described elsewhere[10].

Since only a simple polynomial ratio appears in the WK filter, this is implemented in the discrete time domain as 9 parameters second order A.R.M.A. algorithm applied to raw data sampled at 4882.8125 Hz.

The resulting data at the output of WK filter are very effectively band-limited (see Fig. 2) and can be sub-sampled in order to bury by aliasing the unmodeled features that are present outside the interesting bandwidth. Subsampling is also useful in reducing the data rate. The inverse of the matching filter $M(\omega)$ of eq. ( 14 ) is then applied to the subsampled data, properly translated into the reduced frequency band, thus obtaining the whitened data with PSD $S_w(\omega)$.

The on-line analysis includes a built-in adaptive algorithm that updates the filter parameters to take into account their slow drift on time scales longer than an hour. For example, the core of the algorithm for the estimate of the post-filtering bandwidth (the parameters which mostly affect the SNR) tries to keep $S_w(\omega)$ as flat as possible. This is done for each 2 minute long buffer by comparing the values $S_w(\omega_k)$ averaged on a narrow band around the frequencies of the two modes, with that measured at a selected frequency in between. Since the difference is proportional to the error between the currently used parameters and their optimum value, it is used to drive the adaptive algorithm that adjusts the parameters.

Data in resonant detectors often contain unmodeled signals superimposed to the background gaussian noise. When these signals dominates a stretch of data, the whitening process fails. This is recognized by the adaptive procedure that freezes in the previously adjusted value. This selection procedure allows the filter paramenters to be adjusted for drifts on a time scale longer than the mechanical relaxation time of the system, while ignoring dramatic changes due to isolated events.

When large isolated excitations are present, data are no longer gaussian especially at the high amplitude regions of the distribution. The estimate of noise parameters, in first place $\sigma_A^2$, can then be affected by large biases. In order to ensure self-consistency, the analysis continuously monitors the curtosis of the data and the autocorrelation of the whitened data. If these parameters are found to be within 3 times their expected standard deviations the data buffer is accepted for the filter parameter estimate. Otherwise the filter parameters are frozen in.

If the freezing in of the parameters update occurs too frequently on contiguous data buffer, an alert flag is switched on to indicate instrument malfunctioning. Eventually these flags are the basis for the definition of vetoes on time periods of output data.

A maximum-hold algorithm is applied to the filtered data to search for candidate δ-like GW events. For each event, the time of arrival, the amplitude and $c_a^2$ are estimated. The latter is derived by applying eq. (6) to the sub-sampled whitened data. We use a set $\{y_i\}$ of data long about 3 times the typical WK filter time, $2/\Delta\omega_k^{opt}$, following the event arrival time. This choice ensures that the signal decays into the noise within the selected time span, for signal amplitudes up to SNR=100. The computed $c_a^2$ is attached to the event in the event list.

## 6  Experimental results

A sample of the whitened data taken with the cryogenic detector is shown in Figure 2 within the reduced bandwidth. The flatness of their PSD $S_w(\omega)$ demonstrates the consistency of the model of eq. (6) and the good matching to the parameters of the noise of the detector. The number of degrees of freedom used to compute the $c_a^2$ has been 211, being 212 the number of $\{y_i\}$ samples used to calculate X for this data of the cryogenic detector.

The key result of the present paper is that the estimated $c_a^2$ of each candidate δ-like event does follow the reduced chi-square distribution $c_r^2$, as is shown in Fig. 3 for five days of AURIGA data. In fact, at least at low SNR, the measured $c_a^2$ histograms are well fitted by a chi-square distribution with the proper number of degrees of freedom, as it is expected since most of the events up to SNR=5 are due to statistical fluctuations of the modeled noise. In particular, the estimated amplitude $\hat{A}$ and the $c_a^2$ are indeed independent random variables. The compliance with the chi-square distribution and the independence of $\hat{A}$ and $c_a^2$ are a consequence of two facts: the gaussian nature of the detector noise as a result of the data reduction procedure described above and the consistency of low SNR events with the expected shape of a δ-like mechanical excitation of the antenna.

In order to demonstrate that the WK filter and the chi-square test would correctly recognize a δ-like gravitational wave event, a number of software calibration signals has been numerically added to the real raw data stream acquired over two days from AURIGA. These software signals were given the expected shape to which the WK filter was matched, with SNR of 30 and 45. As Figure 4 shows, the $c_a^2$ of these pulses are in reasonably good agreement with the expected reduced chi-square distribution $c_r^2$ with the proper number of degrees of freedom. A slight distorsion of the observed distribution is accounted for by the fluctuations of the estimate of $\sigma_{\hat{A}}^2$.

To understand the discrimination ability of the test, we show in Figure 5 the Fourier Transform for two high SNR signals taken from the WK filtered real data, one passing the test and the other failing it. The figure shows the remarkable difference in spectral content of the two pulses. It also shows that the shape of the pulse passing the test is in very good agreement with the expected one for a δ-like mechanical excitation of the bar. For the pulse failing the test, the detected pulse is in good agreement with the expected shape for an idealized electromagnetic pulse exciting the SQUID output circuit.

In Figure 6 we show the result of the event search during the normal operation of the AURIGA detector. About 2/3 of the events with SNR>10 can be rejected because they have a $c_a^2$ >1.4, a threshold which corresponds to a confidence level of $1.14 \times 10^{-4}$ for the 211 degrees of freedom we have here. However, only a few per cent of the events with SNR>5 have a $c_a^2$ greater than this rejection threshold of 1.4. So, most of the events complies with the expected shape for an impulsive mechanical excitation of the resonant-bar. Moreover, only about 13% of the events with SNR>5 are accounted for by the modeled noise. We are still investigating on the origin of such a large excess.

In order to assess the validity of the quadratic dependence of computed $c_a^2$ on SNR in Eq. (8), we excited the room temperature resonant-bar detector with electromagnetic pulses applied at the input of the readout amplifier. In Figure 7 we show a scatter plot of the data collected by sending a series of pulses with increasing values of SNR. The plot clearly shows the quadratic dependence of the computed $c_a^2$ on SNR of signals to which the filter is mis-matched. Moreover it shows also that the standard deviation of the computed $c_a^2$ is given to a first approximation by $l \cdot SNR^2$ times the standard deviation of the $\chi_r^2$ distribution with the same number of degrees of freedom. This result holds for SNR high enough to make negligible the contribution of the uncertainty on SNR estimates.

## 7    Discussion and future applications

The reported results clearly shows that at low amplitude the observed $\chi^2$ statistics is in reasonable good agreement with the expected one. At large amplitude the test appears to be able to discriminate between pulses with the expected signal shape and those with a different one. It is worth noticing that for pulses failing the test, a measurement of the value of the parameter $\lambda$ can be used to determine the physical origin of spurious events. For instance, the type of spurious event of the AURIGA detector shown in the lower part of Figure 5 would correspond to a $\lambda \approx 0.01$; therefore the selected threshold of 1.4 on $\chi_r^2$ would efficiently cut spurious events of this type for SNR>7 while leaving unaffected signals with proper shape to a very high confidence level.

It is worth mentioning however that the experimental $c_a^2$ has a probability distribution function slightly distorted in respect to a pure $\chi_r^2$. This is well accounted for by both the need to estimate various noise parameters from the data, a procedure that increases the spread of the distribution, and by the data being non-stationary. The confidence level can be determined empirically by using proper calibration pulses (as for the data in Figure 4), at least for the higher false dismissal probability range.

The value of $\lambda$ for electromagnetic pulses at the SQUID output of the AURIGA detector is smaller of a factor 4 than the Monte-Carlo estimate $\lambda \approx 0.04$ we gave in ref. 4, but this is reasonable taking into account the different setup parameters of the detector used in the simulation. In particular the post-filtering bandwith was $\approx$ 30Hz for the simulation, a much higher value than the presently achieved $\approx$ 1Hz in the detector.

Whatever be the efficiency of the cleaning method described so far in rejecting spurious events, a finite amount of them survive as they are indistinguishable from gravitational wave signals. As a consequence, a single detector can only give an upper limit for the rate of GW events.

Arrays of detectors help overcoming this problem. In a conventional approach, one looks for coincidences among detectors located far apart, that are assumed to be independent. Since the rate of coincidences decreases as a power law with the number of detectors in the array at the exponent, one tries to achieve conditions where the false alarm probability, as evaluated from Poisson statistics, becomes negligibly small.

The maximum likelihood/optimal filtering method however, leads to a somewhat different procedure: one makes a global fit to the data from the N detectors in the array, of some model signal. The quantity to be minimized is then

$$\Lambda(A, t_o, \hat{n}, \psi) =$$

$$= \frac{1}{2} \sum_{\alpha=1}^{N} \sum_{i,k=1}^{M_\alpha} \mu_{ik}^\alpha \left[ x^\alpha(t_i) - As^\alpha(\theta, \phi, \psi) f\left(t_i - t_o - \frac{\vec{r}^\alpha \cdot \hat{n}}{c},\right) \right] \times$$

$$\times \left[ x^\alpha(t_k) - As^\alpha(\theta, \phi, \psi) f\left(t_k - t_o - \frac{\vec{r}^\alpha \cdot \hat{n}}{c}\right) \right] \quad (15)$$

where $\hat{n}$ is the wave unit vector with angles $\theta$ and $\phi$, $\vec{r}^\alpha$ is the position vector of the $\alpha^{th}$ detector in the array with respect to a geocentric coordinate system and $t_o$ is the signal arrival at the center of the Earth. $s^\alpha(\theta,\phi,\psi)$ is a form factor that takes into account that the response of the $\alpha^{th}$ detectors to the *same* incoming wave $Af(t)$ depends on its orientation in respect to the wave vector and on its polarization angle[11] $\psi$.

For each choice of $\hat{n}$, $t_o$ and $\psi$, $\Lambda(A, t_o, \hat{n}, \psi)$ reaches a minimum[1] when A is the weighted average:

$$A_{opt}(t_o, \hat{n}, \psi) = \frac{\sum_{\alpha=1}^{N} \frac{A_{opt}^\alpha(t_o, \hat{n}, \psi)}{\sigma_{A\alpha}^2}}{\sum_{\alpha=1}^{N} \frac{1}{\sigma_{A\alpha}^2}} \quad (16)$$

where $A_{opt}^\alpha(t_o, \hat{n}, \psi)$ is the amplitude estimate *obtained by using the data from the $\alpha^{th}$ detector only.*

One can easily calculate the result that the minimum corresponding chi-square value factorizes according to:

$$\chi^2(t_o, \hat{n}, \psi) \equiv 2\Lambda_{min}(t_o, \hat{n}, \psi) = \sum_{\alpha=1}^{N} \frac{1}{\sigma_\alpha^2} \left[ \sum_{i=1}^{M_\alpha} y_\alpha^2(t_i) \right] - \frac{A_{opt}^2(t_o, \hat{n}, \psi)}{\sigma_A^2} =$$

$$= \sum_{\alpha=1}^{N} \chi_\alpha^2(t_o, \hat{n}, \psi) + \sum_{\alpha=1}^{N} \frac{\left[ A_{opt}^\alpha(t_o, \hat{n}, \psi) - A_{opt}(t_o, \hat{n}, \psi) \right]^2}{\sigma_{A\alpha}^2} \equiv \sum_{\alpha=1}^{N} \chi_\alpha^2(t_o, \hat{n}, \psi) + \chi_g^2(t_o, \hat{n}, \psi) \quad (17)$$

where $\chi_\alpha^2$ is the chi-square one estimates by using the data form the $\alpha^{th}$ detector only and $\chi_g^2$ is the chi-square of the common weighted average $A_{opt}$ of the amplitudes $A_{opt}^\alpha$ estimated by each detector. It can indeed be shown that $\chi_g^2$ is independent of all the $\chi_\alpha^2$'s. This shows that the global chi-square test for an array of detectors can indeed be made by adding the individual chi-square values $\chi_\alpha^2$ for each detector to $\chi_g^2$.

As with any multiple parameter non-linear fit, the procedure should be repeated for all $\hat{n}$ and $\psi$ in search for the absolute minimum. This reintroduces a correlation among the $\chi_\alpha^2$ and $\chi_g^2$ as the global minimum does not coincides with the parameter values that minimize either each $\chi_\alpha^2$ or $\chi_g^2$. Resonant detectors presently in operation[12] are however oriented almost parallel. In addition, full high resolution timing has been implemented up to now only for AURIGA. In practice, due to the still comparatively low bandwidth of these detectors, only phase-timing, i.e. timing modulo a period of antenna oscillation, can be done at reasonable signal-to-noise ratio[8]. As a consequence, each detector produces a list of candidate events with a time of arrival only known within a fraction of a second. Coincidence analysis[13] is then performed with a time window of the same order.

Within this somehow coarse procedure eq. ( 17) still indicates that $\chi_g^2$ can be used as a reference statistics to tests for the consistency of amplitude of a candidate coincidence event. With 5 detectors $\chi_g^2$ is distributed chi-square with 4 degrees of freedom. Application of this test to data from the IGEC[13] detectors is currently under study.

We wish to acknowledge the precious work of the colleagues who helped us in setting up the AURIGA detector, in particular M. Biasotto, E. Cavallini, D. Carlesso, S. Caruso, R. Macchietto and S. Paoli. We are also indebted for helpful discussions with G.V. Pallottino and I.S.Heng. This work has been supported in part by a grant from M.U.R.S.T.-COFIN'97.

**Caption to Figures**

Figure 1: Schematic drawing of the AURIGA data acquisition system. The signal channel from the transducer and d.c.SQUID amplifier system is acquired by the 23 bit ADC at about 4.9 kHz. The synchronization with UTC of the acquired data is achieved in hardware well within 1μs by dedicated interrupts and triggers between the ADC and a GPS clock with a stabilized local oscillator. The full raw data are then fully archived and analysed on-line.

Figure 2: The Power Spectral Density of the raw data around the detector modes (upper) shows only small monochromatic disturbances. The PSD of the WK filtered data (middle) are effectively band-limited and therefore can be suitably subsampled keeping all the information within a 35 Hz bandwidth around the modes. In this bandwidth, the whitened data (lower) demonstrate that the parameters of the noise model were correctly estimated.

Figure 3: Left: plot of $c_a^2$ and SNR of candidate events for AURIGA with 3<SNR<6. Right: histograms of $c_a^2$ of all these events (white) and of events whose SNR is between 3 and 3.5, 3.5 and 4, and so on up to between 5.5 and 6 (from brighter gray to darker gray respectively). The continuous lines are reduced chi-square distributions $c_r^2$ with 211 degrees of freedom fitted to these histograms: the agreement is evident and is independent from the SNR. The data are relative to 5 days of data taking and to about 24000 events above SNR=3.

Figure 4: Left: 3D histogram of $c_a^2$ vs. SNR for AURIGA. Data that cluster around SNR ≈30, SNR≈45 and $c_a^2 \approx 1$ are due to software calibration pulses with shape matched to the WK filter which have been added on the real data stream acquired by AURIGA during 14[th]-15[th] June 1997. Spurious signals are not visible in this range. Right: histograms of $c_a^2$ for the low amplitude candidate events (gray area) and for the software calibration pulses (white and dark gray area).

Figure 5: Fast Fourier Transform of detected candidate events pulses with SNR=18.2 and $c_a^2$ =1.01 (upper) and with SNR=23.5 and $c_a^2$ =6.8 (lower) at the output of the WK filter. The superimposed continuous lines represent the expected responses for a mechanical δ-like excitation of the bar (upper), and a fast electromagnetic excitation entering the ADC input or the SQUID output (lower) respectively. For comparison, the upper continuous line is also shown in the lower graph as a dashed line.

Figure 6: Scatter plot of $c_a^2$ vs. SNR (upper) and SNR histogram of events with $c_a^2$ <1.4 (white area) and for all values of $c_a^2$ (white plus gray). The plots refer to 10 days of candidate events of AURIGA from 12[th] to 21[th] June 1997, corresponding to an effective observation time of 181 hours. The selected threshold of 1.4 used for the $\chi_r^2$ test corresponds to a confidence level for false dismissal of 1.17 x 10$^{-4}$. The test allows to reduce only marginally the number of candidate events with SNR>5, from 1337 to 1306; however, for SNR>10 the $\chi_r^2$ test vetoes about 2/3 of the events. The dashed line in the histogram is the distribution predicted with a simulated quasi-stationary gaussian process, whose post-detection bandwidths follow the same time behavior of the measured ones during the observation time.

It is evident that above SNR=5 the modeled gaussian noise only accounts for about 13% of the detected events.

Figure 7: Plot of computed $\left(c_a^2 - 1\right)$ vs. SNR for spurious electromagnetic impulsive events, using 136 degrees of freedom. These events were generated in the room temperature detector by applying a burst excitation to the input port of the readout amplifier coupled to the motion transducer. The excitation amplitudes are uniformly distributed between SNR=0 and SNR=50. The computed $c_a^2$ distribution follows a quadratic scale law, as in Eq. (8), with $\lambda$=0.029 (thick line). The gray area at low SNR stands for the low SNR background events.

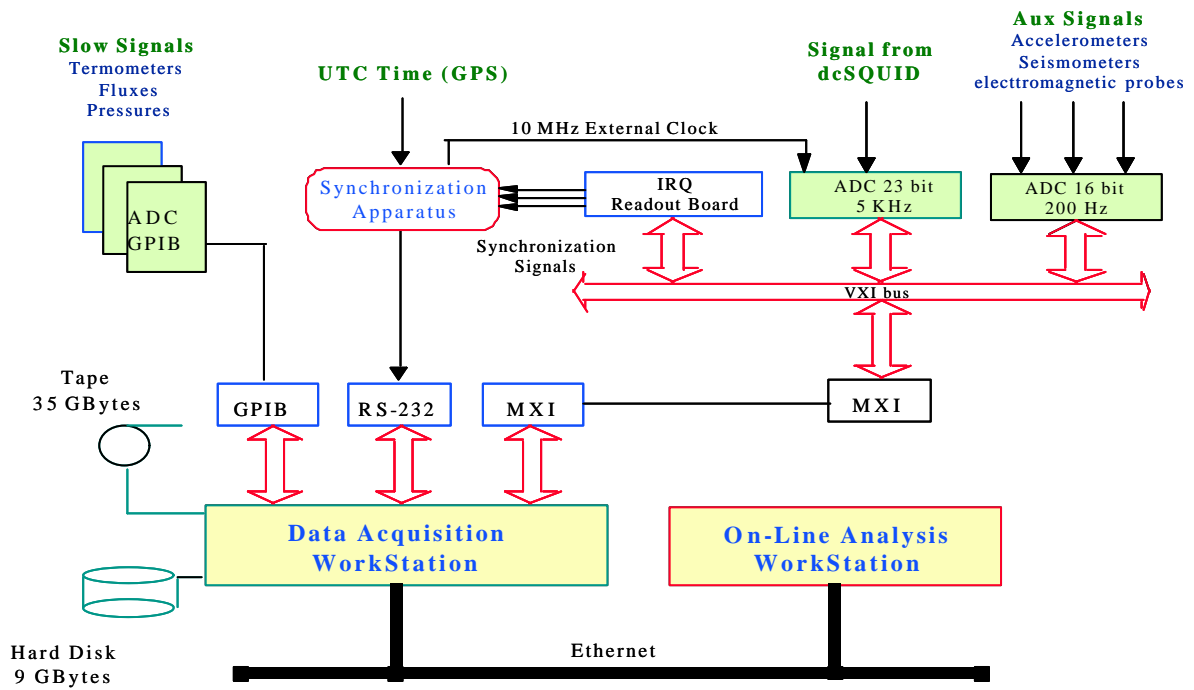

**Fig. 1**

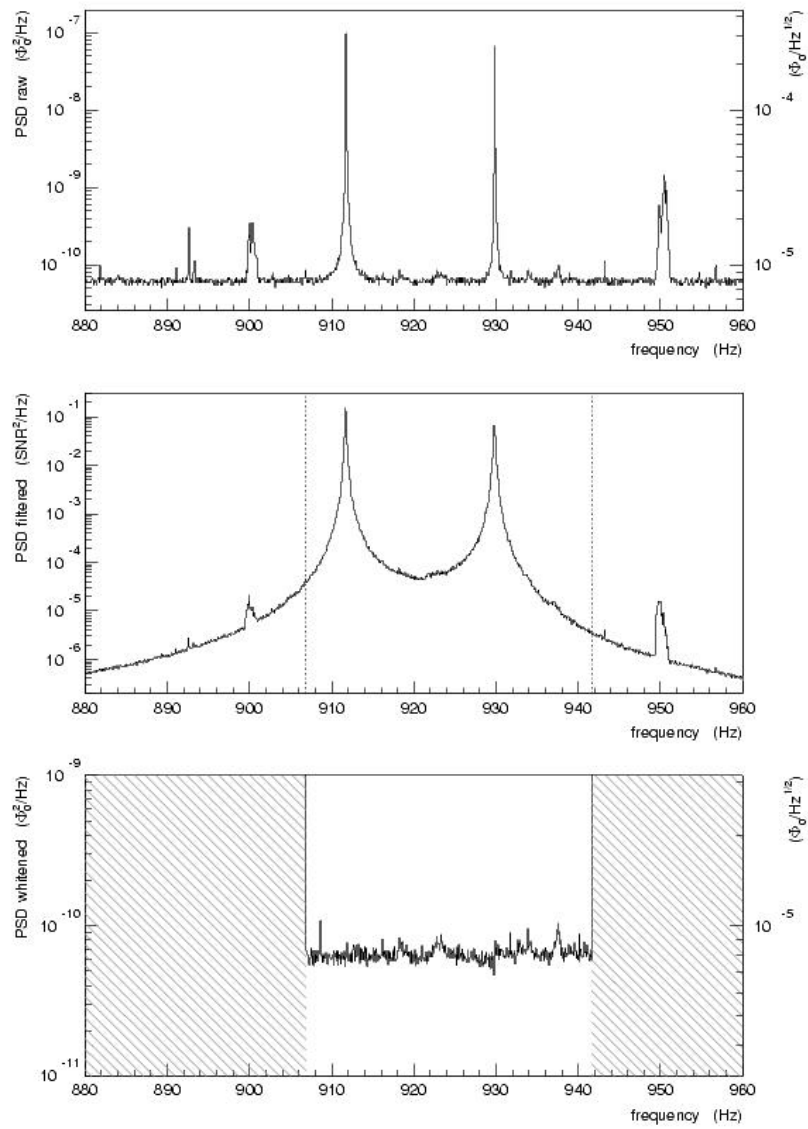

**Fig. 2**

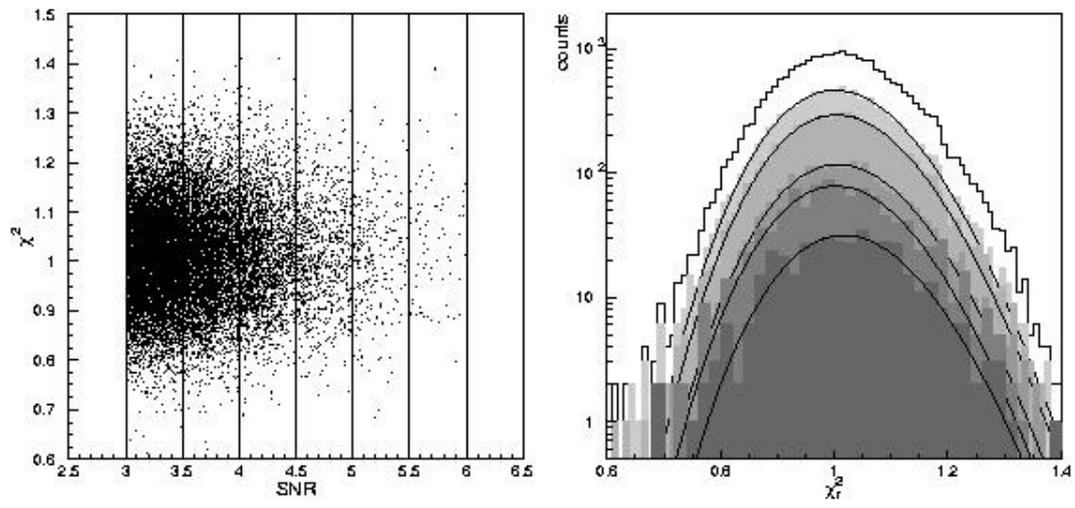

**Fig. 3**

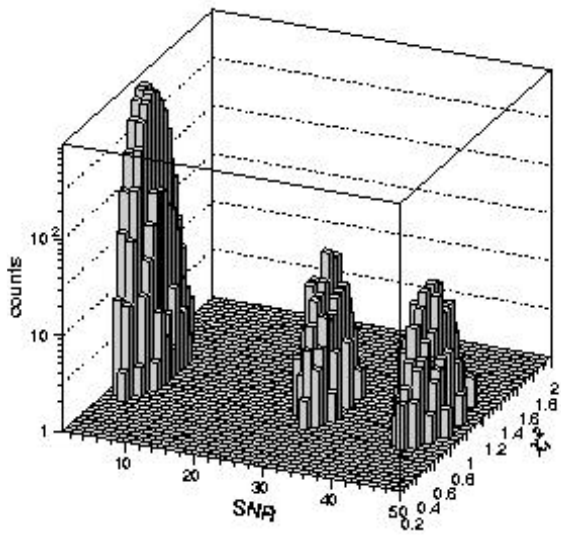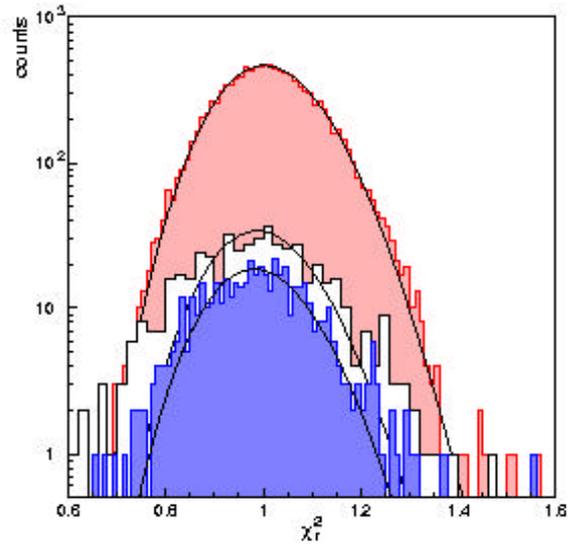

**Fig 4**

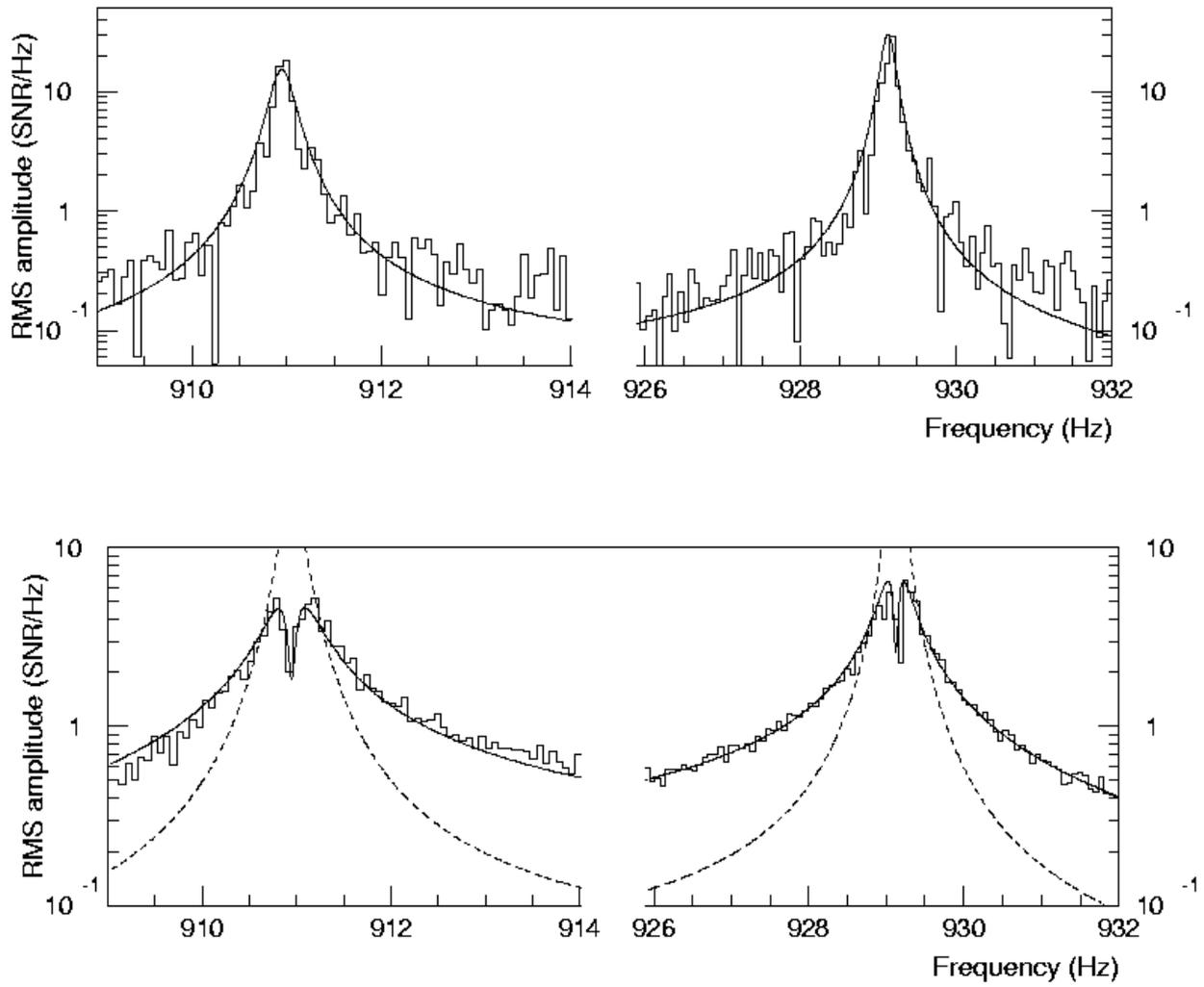

**Fig 5**

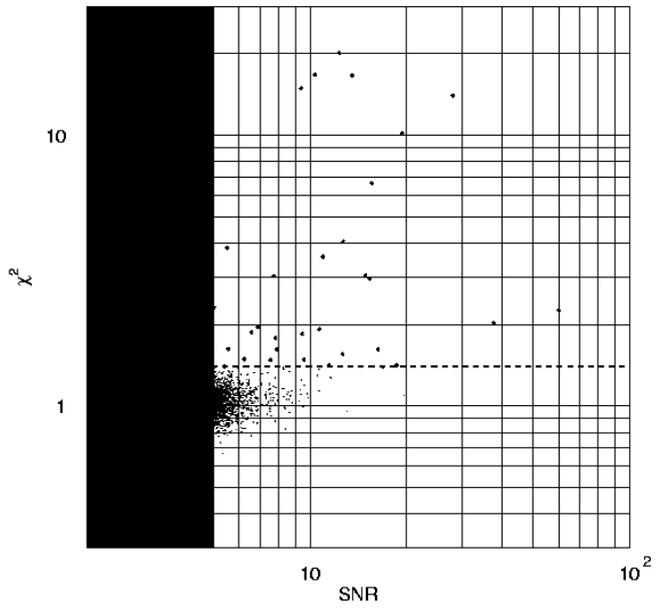

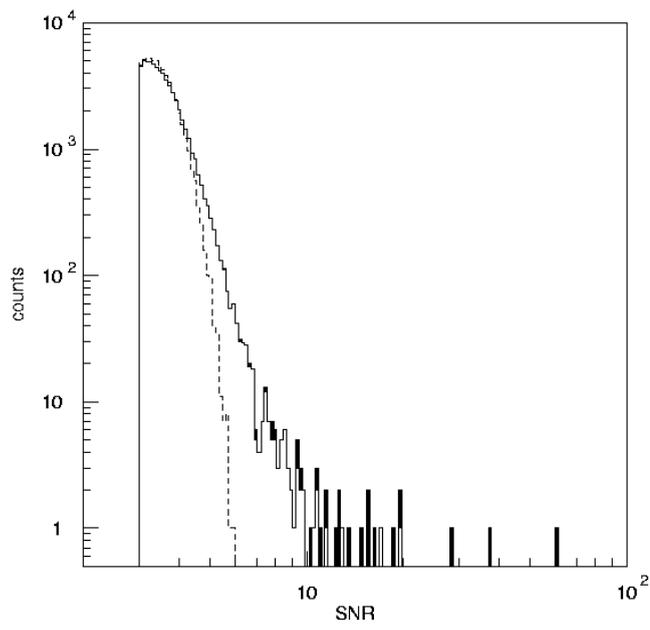

**Fig. 6**

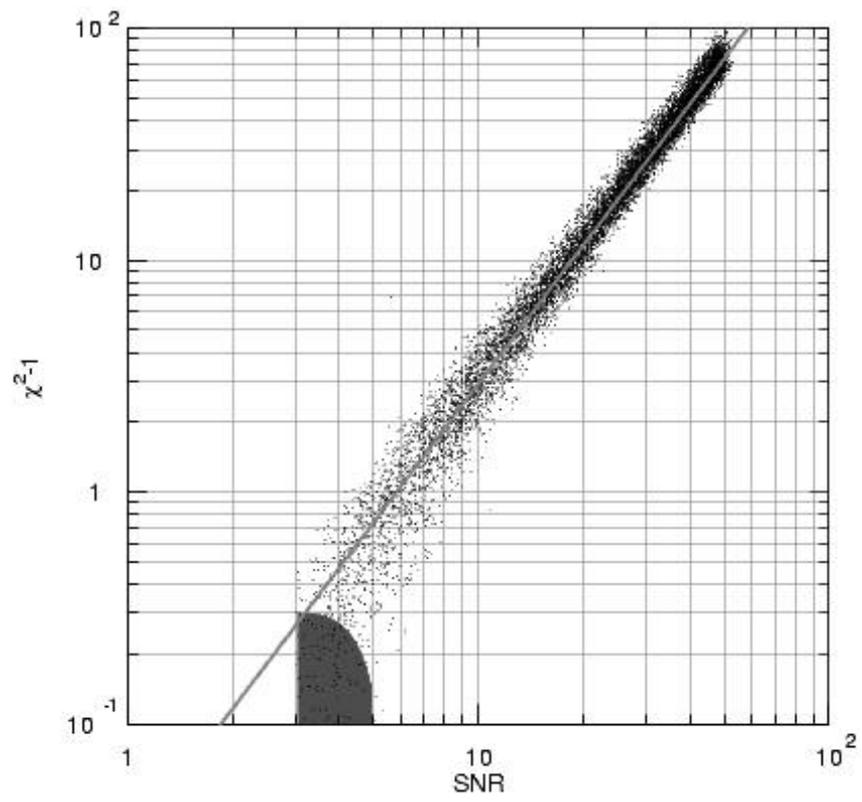

**Fig 7**